\documentclass[a4paper,12pt,english]{article}

\usepackage{helvet}         % selects Helvetica as sans-serif font
\usepackage{courier}        % selects Courier as typewriter font
\usepackage{type1cm}        % activate if the above 3 fonts are
\usepackage{makeidx}         % allows index generation
\usepackage{graphicx}        % standard LaTeX graphics tool
\usepackage{multicol}        % used for the two-column index
\usepackage[bottom]{footmisc}% places footnotes at page bottom

\usepackage{cite}
\usepackage{mathtools}
\usepackage{empheq}

\usepackage[mathscr]{euscript}
\usepackage[all,color]{xy}
\usepackage{relsize}
\usepackage{url}
\usepackage{cancel}
\usepackage{caption}
\usepackage[T1]{fontenc}
\usepackage{color}
\usepackage{mathtools}

\newcommand{\bbm}{\left(\begin{matrix}}
    \newcommand{\ebm}{\end{matrix}\right)}
\newcommand{\beq}{\begin{eqnarray}}
\newcommand{\eeq}{\end{eqnarray}}

\makeatother

 \def\one{\mbox{1 \kern-.59em {\rm l}}}

\begin{document}
\begin{flushright}
CERN-TH-2019-185
\end{flushright}

\begingroup
{\let\newpage\relax% Void the actions of \newpage
\title{Gauge Theories on Fuzzy Spaces and Gravity}

\author{G. Manolakos\textsuperscript{1},\,P. Manousselis\textsuperscript{1},\,G. Zoupanos\textsuperscript{1,2,3}}\date{}
\maketitle}
\begin{center}
\emph{E-mails: gmanol@central.ntua.gr\,, pman@central.ntua.gr\,, George.Zoupanos@cern.ch }
\end{center}

\begin{center}
\itshape\textsuperscript{1}Physics Department, National Technical
University, GR-15780 Athens, Greece\\
\itshape\textsuperscript{2} Theory Department, CERN\\
\itshape\textsuperscript{3} Max-Planck Institut f\"ur Physik, Fohringer Ring 6, D-80805 Munchen, Germany
\end{center}
\vspace{0.1cm} \emph{Keywords}: gauge theories, four-dimensional gravity, noncommutative spaces, fuzzy de Sitter 

\maketitle

\abstract{
We start by briefly reviewing the description of gravity theories as gauge theories in four dimensions.
More specifically we recall the procedure leading to the results of General Relativity and Weyl Gravity in a
gauge-theoretic manner. Then, after a brief reminder of the formulation of gauge theories on noncommutative
spaces, we review our recent work, where gravity is constructed as a gauge theory on the fuzzy $dS_4$.
}

\section{Introduction}
\label{sec:intro}

One of the main research areas addressing the problem of the lack of
knowledge of the spacetime quantum structure is based on the idea that at extremaly small distances (Planck length)
the coordinates exhibit a noncommutative structure.
Then it is natural to wonder which are the implications for gravity of such an idea.
On the other hand at more ordinary (say LHC) distances the Strong, Weak and Electromagnetic interactions are successfully
formulated using gauge theories, while at much smaller distances the Grand Unified Gauge Theories provide a very attractive
unification scheme of the threee interactions.
The gravitational interaction is not part of this picture, admitting a geometric formulation, the Theory of Relativity. However there exists a gauge-theoretic approach to gravity besides the geometric one
\cite{Utiyama:1956sy,Kibble:1961ba,Stelle:1979aj,MacDowell:1977jt,Ivanov:1980tw,Kibble:1985sn,Kaku:1977pa,
Fradkin:1985am,vanproeyen,cham-thesis,Chamseddine:1976bf,Witten:1988hc}.
%[1-12]OK
This approach started with the pioneer work of Utiyama
\cite{Utiyama:1956sy}
%[1]OK
and was refined by other authors
\cite{Kibble:1961ba,Stelle:1979aj,MacDowell:1977jt,Ivanov:1980tw,Kibble:1985sn,Kaku:1977pa,
Fradkin:1985am,vanproeyen,cham-thesis,Chamseddine:1976bf,Witten:1988hc}
%[2-12]OK
as a gauge theory of the de Sitter $SO(1,4)$ group, spontaneously broken by a scalar field to the Lorentz $SO(1,3)$ group.
Similarly using the gauge-theoretic approach the Weyl gravity has been constructed as a gauge theory of the 4-d conformal group
\cite{Kaku:1977pa,Fradkin:1985am}.
%[7,8]OK
Returning to the noncommutative framework and taking into account the gauge-theoretic description of gravity,
the well-established formulation of gauge theories on noncommutative spaces leads to the construction of models
of noncommutative gravity
\cite{Chamseddine:2000si,Chamseddine:2003we,Aschieri1,Aschieri2,Ciric:2016isg,Cacciatori:2002gq,
Cacciatori:2002ib,Aschieri3,Banados:2001xw}.
%[37-45]OK
In these treatments the authors use the constant noncommutativity (Moyal-Weyl), the formulation of the $\star$-product and
the Seiberg-Witten map
\cite{Seiberg:1999vs}.
%[46]OK
In addition to these treatments noncommutative gravitational models can be constructed using the noncommutative
realization of matrix geometries
\cite{Banks:1996vh,Ishibashi:1996xs,Aoki:1998vn,Hanada:2005vr,Furuta:2006kk,Yang:2006dk,Steinacker:2010rh,Kim:2011cr,Nishimura:2012xs,Nair:2001kr,Abe:2002in,Valtancoli:2003ve,Nair:2006qg},
%[47-59]OK
while it should also be noted that there exist alternative approaches
\cite{Buric:2006di,Buric:2007zx,Buric:2007hb}
%[60-62]OK
(see also \cite{Aschieri:2003vyAschieri:2004vhAschieri:2005wm}),
%[63]OK
which will not be considered here.
It should also be noted that the formulation of noncommutative gravity implies, in general, noncommutative
deformations which break the Lorentz invariance. However, ``covariant noncommutative spaces'' have been constructed
too
\cite{Yang:1947ud,Heckman:2014xha}
%[64-65]OK
which preserve the Lorentz invariance. Consequently noncommutative deformations of field theories have been constructed 
\cite{Chatzistavrakidis:2018vfi,Manolakos:2018isw,Manolakos:2018hvn,Hammou:2001cc,Vitale:2014hca,Kovacik:2013yca,Jurman:2013ota,Manolakos:2019fle,Chamseddine:2002fd,Li:1973mq}
(see also
\cite{Buric:2015wta,Sperling:2017dts,Kimura:2002nq,Buric:2017yes,Steinacker:2016vgf}).
%[67-70]OK
The main point of this article is to present the various features of a 4-d gravity that we have constructed recently
\cite{Manolakos:2019fle}
%[78-80]OK
as a gauge theory on a fuzzy $dS_4$. Motivated by Heckman-Verlinde
\cite{Heckman:2014xha},
%[66]OK
who were based on Yang's early work
\cite{Yang:1947ud},
%[65]OK
we have considered a 4-d covariant fuzzy $dS$ space which preserves Lorentz invariance. The requirement
of covariance led us to an enlargement of the isometries of the fuzzy $dS_4$, specifically from $SO(1,4)$ to $SO(1,5)$.
Then the construction of a gauge theory on this noncommutative space by gauging a subgroup of the full isometry,
led us to an enlargement of the gauge group and in fixing its representation. In addition the covariance of the
field strength tensor required the inclusion of a 2-form gauge field. Eventually we have proposed an action of Yang-Mills type,
including the kinetic term of the 2-form.

\section{Gravity as a gauge theory}
\label{sec:GR-GAUGE}

In this section we recall the interpretation of the four-dimensional Einstein and Weyl gravities
as gauge theories in order to be used later in the framework of noncommutative fuzzy spaces.

\subsection{4-d Einstein's Gravity as a Gauge Theory}
\label{subsec:4-d-Einstein}

Gravitational interaction in four dimensions is described by General Relativity, a solid and successful
theory which has been well-tested over decades since its early days. It is formulated geometrically in contrast
to the rest of the interactions, which are described as gauge theories.
Targeting to a unified description of gravity with the other interactions, a gauge-theoretic approach to gravity
has been developed
\cite{Utiyama:1956sy,Kibble:1961ba,Stelle:1979aj,MacDowell:1977jt,Ivanov:1980tw,Kibble:1985sn}.
%[1-6].
Lets us recall the main features of this approach to describe the 4-d Einstein's gravity.
To achieve a gauge-theoretic approach of 4-d gravity, as a first step the vierbein formulation of General Relativity
has to be employed.
Then depending on the presence and sign of the cosmological constant gauge theories have been constructed on the
Minkowski $M^4$, de Sitter $dS_4$ and anti-de-Sitter $AdS_4$ spacetimes based on the gauge groups Poincare, de Sitter
and Anti-de Sitter, respectively.
The choice of these groups as the symmetry gauge groups being that they are the isometry groups of the
corresponding spacetimes.
Let us start with the case in which there is no cosmological constant
included, i.e., the case of the Poincar\'e group. In this case the
generators of the corresponding algebra satisfy the following commutation relations:
\begin{equation}%2.11
[M_{ab},M_{cd}]=4\eta_{[a[c}M_{d]b]}~,\,\,\,\,\,
[P_a,M_{bc}]=2\eta_{a[b}P_{c]}~,\,\,\,\,\,
[P_a,P_b]=0~,
\end{equation}
where $\eta_{ab} = diag (-1, 1, 1, 1)$ is the metric tensor of the 4-d Minkowski spacetime, $M_{ab}$ are the generators
of the Lorentz group (the Lorentz transformations) and $P_a$ are the generators of the local translations.
Then according to the standard gauging procedure, the gauge potential, $A_\mu$, is introduced and it is expressed
as a decomposition on the generators of the Poincare algebra, as follows:
\begin{equation}%2.12
\label{connection}
A_{\mu}(x)=e_{\mu}{}^a(x)P_a+\frac{1}{2}\omega_{\mu}{}^{ab}(x)M_{ab}~.
\end{equation}
The functions attached to the generators are the gauge fields of the theory and, in this case, they are identified
as the vierbein, $e_{\mu}{}^a$ , and the spin connection, $\omega_{\mu}{}^{ab}$ , which correspond to the translations,
$P_a$, and the Lorentz generators, $M_{ab}$, respectively. In this way, i.e. considering the vierbein as gauge field,
it is achieved a mixing among the internal and spacetime symmetries and that is what makes this kind
of construction special, as compared to the gauge theories describing other interactions. The gauge connection
$A_\mu$ transforms according to the following rule:
\begin{equation}%2.13
\label{transformconnection}
\delta A_{\mu}=\partial_{\mu}\epsilon+[A_{\mu},\epsilon]~,
\end{equation}
where $\epsilon = \epsilon (x)$ is the gauge transformation parameter which is also expanded on the generators of the algebra:
\begin{equation}%2.14
\label{parameter}
\epsilon(x)=\xi^a(x) P_a+\frac{1}{2} \lambda^{ab}(x)M_{ab}~.
\end{equation}
Combining eqs (\ref{connection}) and (\ref{parameter}) with (\ref{transformconnection}) result to the following expressions
of the transformations of the
gauge fields:
\begin{align}%2.15, 2.16
\delta e_{\mu}{}^a&=\partial_{\mu}\xi^a+\omega_{\mu}{}^{ab}\xi_b-\lambda^a{}_be_{\mu}{}^b~,
\\
\delta \omega_{\mu}{}^{ab}&=\partial_{\mu}\lambda^{ab}-2\lambda^{[a}{}_c\omega_{\mu}{}^{cb]}~.
\end{align}
According to the standard procedure followed in gauge theories, the corresponding field strength tensor of the
gauge theory is defined as:
\begin{equation}%2.17
\label{usualformula}
R_{\mu\nu}(A)=2\partial_{[\mu}A_{\nu]}+[A_{\mu},A_{\nu}]~
\end{equation}
and since it is valued in the algebra of generators is also expanded on them as:
\begin{equation}
\label{expansion_of_curvature}%2.18
R_{\mu\nu}(A)=R_{\mu\nu}{}^a(e)P_a+\frac{1}{2} R_{\mu\nu}{}^{ab}(\omega)M_{ab}~,
\end{equation}
where $R_{\mu\nu}{}^a$ and $R_{\mu\nu}{}^{ab}$ are the curvatures associated to the component gauge fields, identified as
the torsion and curvature, respectively. Replacing eqs (\ref{connection}) and (\ref{expansion_of_curvature})
in the (\ref{usualformula}) results to the following
explicit expressions:
\begin{align}%2.19, 2.20
R_{\mu\nu}{}^a(e)&=2\partial_{[\mu}e_{\nu]}{}^a-2\omega_{[\mu}{}^{ab}e_{\nu]b}~,\\
R_{\mu\nu}{}^{ab}(\omega)&=2\partial_{[\mu}\omega_{\nu]}{}^{ab}-2\omega_{[\mu}{}^{ac}\omega_{\nu]c}{}^b~.\label{curvtwoform}
\end{align}
Concerning the dynamics of the theory, the obvious choice is an action of Yang-Mills type, invariant under the
gauge Poincar\'e group ISO(1,3).
However, the aim is to result with the Einstein-Hilbert action, which is Lorentz invariant and, therefore, the
gauge Poincar\'e group ISO(1,3) of
the initial action has to be broken to the gauge Lorentz group SO(1,3).
This can be achieved by  gauging the SO(1,4) group, instead of the Poincar\'e group ISO(1,3),
and employing its spontaneous symmetry breaking, induced by a scalar field that belongs to its fundamental
representation
\cite{Stelle:1979aj,Ivanov:1980tw}.
%[3,5].
The choice of the 4-d de Sitter group is an alternative and preferred choice to that of the Poincare group,
since all generators of the algebra can be considered on equal footing. The spontaneous symmetry breaking leads
to the breaking of the translational generators, resulting to a constrained theory with vanishing torsion involving
the Ricci scalar (and a topological Gauss-Bonnet term), respecting only the Lorentz symmetry, that is the
Einstein-Hilbert action!

Concluding, Einstein's four-dimensional gravity can be formulated as a gauge theory of the Poincare group,
as far as the kinematic part is concerned, i.e. the transformation of the fields and the expressions of the
curvature tensors. Going to the dynamics though, instead of the Poincare group, it is the de Sitter symmetry
which the initial Yang-Mills action has to respect. In turn, the inclusion of a scalar field and the addition
of an appropriate kinetic term in the Lagrangian leads to a spontaneous symmetry breaking to the Lorentz gauge
symmetry, i.e. to the Einstein-Hilbert action.

An alternative way to obtain an action with Lorentz symmetry, is to impose that the action is invariant only
under the Lorentz symmetry and not under the total Poincare symmetry with which one starts. This means that
the curvature tensor related to the translations has to vanish. In other words the torsionless condition is
imposed in this way as a constraint that is necessary in order to result with an action respecting only the
Lorentz symmetry. Solution of this constraint leads to a relation of the spin connection with the vielbein:
\begin{equation}%2.21
\label{spin-viel}
\begin{split}
\omega_\mu^{~ab}&=\frac{1}{2}e^{\nu a}(\partial_\mu e_\nu^{~b}-\partial_\nu e_\mu^{~b})
\frac{1}{2}e^{\nu b}(\partial_\mu e_\nu^{~a}-\partial_\nu e_\mu^{~a})\\
&\qquad\qquad -\frac{1}{2}e^{\rho a}e^{\sigma b}(\partial_\rho e_{\sigma c}-\partial_\sigma e_{\rho c})e_\mu^{~c}\,.
\end{split}
\end{equation}

However, straightforward consideration of an action of Yang-Mills type with Lorentz symmetry, would lead to
an action involving the $R(M)^2$ term, which is not the correct one, since the aim is to obtain the Einstein-Hilbert action.
Also, such an action would imply the wrong dimensionality (zero) of the coupling constant of gravity.
In order to result with the Einstein-Hilbert action, which includes a dimensionful coupling constant,
the action has to be considered in an alternative, non-straightforward way, that is the construction of
Lorentz invariants out of the quantities (curvature tensor) of the theory. The one that is built by
certain contractions of the curvature tensor is the correct one, ensuring the correct dimensionality
of the coupling constant, and is identified as the Ricci scalar and the corresponding action is eventually
the Einstein-Hilbert action.

\subsection{4-d Weyl Gravity as a Gauge Theory}
Besides Einstein's gravity, also Weyl's gravity has been successfully described as a gauge theory of the 4-d conformal group, SO(2,4).
In this case, too, the transformations of the fields and the expressions of the curvature tensors are determined in a straightforward way.
The initial action that is considered is an SO(2,4) gauge invariant action of Yang-Mills type which is broken by imposition of specific
conditions (constraints) on the curvature tensors.  After taking into account the constraints, the resulting action of the theory is
the scale invariant Weyl action
\cite{Kaku:1977pa,Fradkin:1985am,vanproeyen}
%[7-9]
(see also
\cite{cham-thesis,Chamseddine:1976bf}).
%[10,11].

The generators of the conformal algebra of SO(2,4) are the local translations ($P_a$), the Lorentz transformations ($M_{ab}$), the conformal
boosts ($K_a$) and the dilatations ($D$). Their algebra is determined by
their commutation relations:
\begin{equation} %2.22
\begin{split}
[M_{ab},M^{cd}]&=4M_{[a}^{~[d}\delta_{b]}^{c]}\,,\quad [M_{ab},P_c]=2P_{[a}\delta_{b]c}\,,\quad [M_{ab},K_c]=2K_{[a}\delta_{b]c}\,\\
[P_a,D]&=P_a\,,\quad [K_a,D]=-K_a\,,\quad [P_a,K_b]=2(\delta_{ab}D-M_{ab})\,,
\end{split}
\end{equation}
where $a, b, c, d = 1 . . . 4$. Then, according to the gauging procedure, the gauge potential, $A_\mu$ of the theory is in turn determined and is given as
an expansion on the generators of the gauge group, i.e.:
\begin{equation}%2.23
\label{connection-conformal}
A_{\mu} = e_{\mu}^{~a}P_{a} + \frac{1}{2}\omega_{\mu}^{~ab}M_{ab} + b_{\mu}D + f_{\mu}^{~a}K_{a}\,,
\end{equation}
where a gauge field has been associated with each generator. In this case, too, the vierbein and the spin connection are identified as gauge fields of the theory.
The transformation rule of the gauge potential, (\ref{connection-conformal}), is given by:
\begin{equation}%2.24
\label{commrule}
\delta_{\epsilon}A_{\mu} = D_{\mu}\epsilon = \partial_{\mu}\epsilon+ [A_{\mu},\epsilon]\,,
\end{equation}
where $\epsilon$ is a gauge transformation parameter valued in the Lie algebra of the SO(2,4) group and therefore it can be written as:
\begin{equation}%2.25
\label{parameter-conformal}
\epsilon = \epsilon_{P}^{~a} P_{a} + \frac{1}{2}\epsilon_{M}^{~~ab} M_{ab} + \epsilon_{D}D+ \epsilon_{K}^{~a}K_{a}\,.
\end{equation}
Combining the equations (\ref{commrule}), (\ref{connection-conformal}) and (\ref{parameter-conformal}) result to the following expressions of the transformations
of the gauge fields of the theory:
\begin{equation}
\label{conformaltrans}
\begin{split}%2.26
\delta e_{\mu}^{~a} &= \partial_\mu\epsilon_P^{~a}+2ie_{\mu b}\epsilon_M^{~ab}-i\omega_\mu^{~ab}\epsilon_{Pb}-b_\mu\epsilon_K^{~a}+f_\mu^{~a}\epsilon_D\, , \\
\delta \omega_{\mu}^{~ab} &= \frac{1}{2}\partial_\mu\epsilon_M^{~ab}+4ie_\mu^{~a}\epsilon_P^{~b}+\frac{i}{4}\omega_\mu^{~ac}\epsilon_{M~c}^{~~b}+if_\mu^{~a}\epsilon_K^{~b}\,,  \\
\delta b_{\mu} &= \partial_\mu\epsilon_D-e_\mu^{~a}\epsilon_{Ka}+f_\mu^{~a}\epsilon_{Pa}\,,  \\
\delta f_{\mu}^{~a} &= \partial_\mu\epsilon_K^{~a}+4ie_\mu^{~a}\epsilon_D-i\omega_\mu^{~ab}\epsilon_{Kb}-4ib_\mu\epsilon_P^{~a}+if_\mu^{~b}\epsilon_{M~b}^{~~a}\,.
\end{split}
\end{equation}
Accordingly the field strength tensor is defined by the relation:
\begin{equation}%2.27
\label{fieldstrengthconformal}
R_{\mu\nu} = 2 \partial_{[\mu} A_{\nu]} - i [ A_{\mu} ,A_{\nu} ]
\end{equation}
and is expanded on the generators as:
\begin{equation}%2.28
\label{conformalexpansion}
R_{\mu\nu}=\tilde{R}_{\mu\nu}^{~~a}P_a+\frac{1}{2}R_{\mu\nu}^{~~ab}M_{ab}+R_{\mu\nu}+R_{\mu\nu}^{~~a}K_a~.
\end{equation}
Then combining the equation (\ref{fieldstrengthconformal}) and (\ref{conformalexpansion}) result in the following expressions of the component curvature tensors:
\begin{equation}%2.29
\label{conformalcurvatures}
\begin{split}
R_{\mu\nu}^{~~~a}(P) &= 2 \partial_{[\mu}e_{\nu]}^{~~a} + f_{[\mu}^{~~a}b_{\nu]} +  e^{~~b}_{[\mu} \omega_{\nu]}^{~~ac} \delta_{bc}, \\
R_{\mu\nu}^{~~~ab}(M) &= \partial_{[\mu} \omega_{\nu]}^{~~ab} + \omega_{[\mu}^{~~ca} \omega_{\nu]}^{~~db} \delta_{cd} + e_{[\mu}^{~~a}e_{\nu]}^{~~b} + f_{[\mu}^{~~a}f_{\nu]}^{~~b},  \\
R_{\mu\nu}(D) &= 2 \partial_{[\mu}b_{\nu]} + f_{[\mu}^{~~a}e_{\nu]}^{~~b}\delta_{ab}, \\
R_{\mu\nu}^{~~~a}(K) &= 2 \partial_{[\mu}f_{\nu]}^{~~a} + e_{[\mu}^{~~a}b_{\nu]} + f_{[\mu}^{~~b}\omega_{\nu]}^{~~ac}\delta_{bc}\,.
\end{split}
\end{equation}

Concerning the action, it is taken to be a gauge SO(2,4) invariant of Yang-Mills type. Then the initial SO(2,4) gauge symmetry can be broken by
the imposition of certain constraints
\cite{{Kaku:1977pa,Fradkin:1985am,vanproeyen}},
%[7-9],
namely the torsionless condition, $R(P) = 0$ and an additional constraint on $R(M)$. The two constraints admit
an algebraic solution leading to expressions of the fields $\omega_\mu^{~~ab}$ and $f_\mu^{~~a}$ in terms of the independent fields
$e_\mu^{~~a}$
and $b_\mu$. In addition, $b_\mu$ can be gauged fixed to $b_\mu$ = 0 and, imposing
all the constraints in the initial action lead to the well-known Weyl action, which is diffeomorphism and scale invariant.

Besides the above breaking of the conformal symmetry which led to the Weyl action, another breaking pattern via constraints has been suggested
\cite{Chamseddine:2002fd},
%[81],
leading to an action with Lorentz symmetry, i.e. explicitly the Einstein-Hilbert action. From our prespective, the latter can be achieved
through an alternative symmetry breaking mechanism, specifically with the inclusion of two scalar fields in the fundamental representation of the
conformal group
\cite{Li:1973mq}.
%[80].
Then the spontaneous symmetry breaking could be triggered just as a generalization of the case of the breaking of the 4-d de Sitter group down to the Lorentz
group by the inclusion of a scalar in the fundamental representation of SO(1,4), as discussed in section \ref{subsec:4-d-Einstein}.
Calculations and details on this issue will be included in a future work.

Moreover, the argument used in the previous section in the 4-d Poincar\'e gravity case as an alternative way to break the initial symmetry to the Lorentz,
can be generalized in the case of conformal gravity too. Since it is desired to result with the Lorentz symmetry starting from the initial gauge SO(2,4)
symmetry, the vacuum of the theory is considered to be directly SO(4) invariant, which means that every other tensor, except for the R(M), has to vanish.
Setting these tensors to zero will produce the constraints of the theory leading to expressions that relate the gauge fields. In particular, in
\cite{Chamseddine:2002fd},
%[81],
it is argued that if both tensors $R(P)$ and $R(K)$ are simultaneously set to zero, then from the constraints of the theory it is understood that the corresponding
gauge fields, $f_\mu^{~~a}$, $e_\mu^{~~a}$ are equal - up to a rescaling factor - and $b_\mu = 0$.

\section{Gauge Theories on Noncommutative Spaces}
Let us now briefy recall the main concepts of the formulation of gauge theories on noncommutative spaces, in order to use them later in the
construction of the noncommutative gravity models.

Gauge fields arise in noncommutative geometry and in particular on fuzzy spaces very naturally; they are linked to the notion of covariant coordinate
\cite{Madore:2000en}.
%[36].
Consider a field $\phi(X_a )$ on a fuzzy space described by the non-commuting coordinates $X_a$ and transforming according to a gauge group $G$. An infinitesimal
gauge transformation $\delta\phi$ of the field $\phi$ with gauge transformation parameter $\lambda(X_a)$ is defined by:
\begin{equation}%3.1
\label{gaugetransffuzzy}
  \delta\phi(X)=\lambda(X)\phi(X)\,.
\end{equation}
If $\lambda (X )$ is a function of the coordinates, $X_a$, then it is an infinitesimal Abelian transformation and $G = U(1)$, while if $\lambda (X )$ is valued in the Lie algebra
of hermitian $P\times P$ matrices, then the transformation is non-Abelian and the gauge group is $G = U(P)$.
The coordinates are invariant under an infinitesimal transformation of the the gauge group, $G$, i.e. $\delta X_a = 0$.
In turn the gauge transformation of the product of a coordinate and the field is not covariant:
\begin{equation}%3.2
  \delta(X_a\phi)=X_a\lambda(X)\phi\,,
\end{equation}
since, in general, it holds:
\begin{equation}%3.3
  X_a\lambda(X)\phi\neq\lambda(X)X_a\phi\,.
\end{equation}
Following the ideas of the construction of ordinary gauge theories, where a covariant derivative is defined, in the noncommutative case,
the covariant coordinate, $\phi_a$, is introduced by its transformation property:
\begin{equation}%3.4
  \delta(\phi_a\phi)=\lambda\phi_a\phi\,,
\end{equation}
which is satisfied if:
\begin{equation}%3.5
\label{transformationofphia}
  \delta(\phi_a)=[\lambda,\phi_a]\,.
\end{equation}
Eventually, the covariant coordinate is defined as:
\begin{equation}%3.6
\label{covariantfield}
  \phi_a\equiv X_a+A_a\,,
\end{equation}
where $A_a$ is identified as the gauge connection of the noncommutative gauge theory. Combining equations (\ref{transformationofphia}), (\ref{covariantfield}), the gauge transformation
of the connection, $A_a$ , is obtained:
\begin{equation}%3.7
  \delta A_a=-[X_a,\lambda]+[\lambda,A_a]\,.
\end{equation}
justifying the interpratation of $A_a$ as a gauge field
\footnote{For more details see \cite{Aschieri:2003vyAschieri:2004vhAschieri:2005wm}}. %[63]}
Correspondingly the field strength tensor, $F_{ab}$ , is defined as:
\begin{equation}%3.8
  F_{ab}\equiv[X_a,A_b]-[X_b,A_a]+[A_a,A_b]-C^c_{ab}A_c=[\phi_a,\phi_b]-C^c_{ab}\phi_c\,,
\end{equation}
which is covariant under a gauge transformation,
\begin{equation}%3.9
  \delta F_{ab}=[\lambda,F_{ab}]\,.
\end{equation}
In the following sections, the above methodology will be applied in the construction of gravity models as gauge theories on fuzzy spaces.

\section{A 4-d Noncommutative Gravity Model}

Let us now proceed with the presentation of a 4-d gravity model as a gauge theory on a fuzzy space.
We start with the construction of an appropriate 4-d fuzzy space and then we build a gravity theory
as a gauge theory on this noncommutative space.

\subsection{Fuzzy de Sitter Space}
Let us construct first the fuzzy 4-d de Sitter space, $dS_4$, which will be used as the background
space on which we will define the gauge theory that we propose to describe gravity. The continuous
$dS_4$ is defined as a submanifold of the 5-d Minkowski spacetime and can be viewed
as the Lorentzian analogue of the definition of the four-sphere as an embedding in the 5-d Euclidean space.
The defining embedding equation of $dS_4$ is:
\begin{equation}%5.1
\eta^{MN}x_Mx_N=R^2\,,
\end{equation}
$M,N=0,\ldots, 4$ and $\eta^{MN}$ is the metric tensor of the 5-d Minkowski spacetime, $\eta^{MN}=\mathrm{diag}(-1,+1,+1,+1,+1)$.
In order to obtain the fuzzy analogue of this space, one has to consider its coordinates, $X_m$, to be operators
that do not commute with each other:
\begin{equation}%5.2
\label{noncommgeneral}
[X_m,X_n]=i\theta_{mn}\,,
\end{equation}
where the spacetime indices are $m,n=1,\ldots,4$. In analogy to the fuzzy sphere case, where the corresponding coordinates
are identified as the rescaled three generators of $SU(2)$ in a high N-dimensional representation, we expect that the
right hand side in eq(\ref{noncommgeneral}), should be identified with a generator of the underlying algebra, ensuring covariance,
i.e $\theta_{mn}=C_{mn}^{~~~r}X_r$, where $C_{mnr}$ is a rescaled Levi-Civita symbol. Otherwise, if the the right hand side in
eq(\ref{noncommgeneral}) is a fixed antisymmetric tensor the Lorentz invariance will be violated. However,
in the present fuzzy de Sitter case, such an identification cannot be achieved, since the algebra is not closing
\cite{Heckman:2014xha}
%[42 sti nea arithmisi] *.
\footnote{
For more details on this issue, see \cite{Sperling:2017dts,Kimura:2002nq},
%53 nea, 86 palia],
where the same problem emerges in the construction of the fuzzy four-sphere.}.
To achieve covariance, the suggestion
\cite{Yang:1947ud,Heckman:2014xha}
%[41,42]
is to use a group with a larger symmetry, in which we will be able to incorporate all generators and the
noncommutativity in it. The minimal extension of the symmetry leads us to adopt the $SO(1,5)$ group. Therefore,
a fuzzy $dS_4$ space, with its coordinates being operators represented by N-dimensional matrices, respecting
covariance, too, is obtained after the enlargement of the symmetry to the $SO(1,5)$
\cite{Manolakos:2019fle}.
%[49 sti nea arithmisi]].
To facilitate the construction we make use of the Euclidean signature, therefore, instead of the $SO(1,5)$,
the resulting symmetry group is considered to be that of $SO(6)$.

In order to formulate explicitly the above 4-d fuzzy space, let us consider the $SO(6)$ generators,
denoted as $\mathrm{J}_{AB} = - \mathrm{J}_{BA}$, with $A,B = 1,\ldots, 6$, satisfying the following commutation relation:
\begin{equation}%5.3
[J_{AB}, J_{CD} ] = i(\delta_{AC}J_{BD} + \delta_{BD}J_{AC} - \delta_{BC}J_{AD} - \delta_{AD}J_{BC} )\,.
\end{equation}
These generators can be written as a decomposition in an $SO(4)$ notation, with the component generators
identified as various operators, including the coordinates, i.e.:
\begin{equation}%5.4
J_{mn} = \tfrac{1}{\hbar} \Theta_{mn}, \ \ J_{m5} = \tfrac{1}{\lambda} X_{m}, \ \  J_{m6} = \tfrac{\lambda}{2 \hbar}P_{m} , \ \  J_{56} = \tfrac{1}{2} \mathrm{h}\,,
\end{equation}
where $m,n = 1,\ldots,4$. For dimensional reasons, an elementary length, $\lambda$, has been introduced in the above
identifications, in which the coordinates, momenta and noncommutativity tensor are denoted as $X_{m}$, $P_{m}$ and $\Theta_{mn}$,
respectively. Then the coordinate and momentum operators satisfy the following commutation relations:
\begin{align}
[ X_{m} , X_{n} ] = i \frac{\lambda^{2}}{\hbar} \Theta_{mn}, &\qquad [P_{m}, P_{n} ] = 4i \frac{\hbar}{\lambda^{2}} \Theta_{mn},
\\
[ X_{m}, P_{n} ]  = i \hbar \delta_{mn}\mathrm{h}, &\qquad [X_{m}, \mathrm{h} ] = i \frac{\lambda^{2}}{\hbar} P_{m},
\\
[P_{m}, \mathrm{h} ]& =4i \frac{\hbar}{\lambda^{2}} X_{m}\,,
\end{align}
%5.5, 5.6, 5.7
while the algebra of spacetime transformations is given by:
\begin{align}
[X_{m}, \Theta_{np} ] &= i \hbar ( \delta_{mp} X_{n} - \delta_{mn} X_{p} )
\\
[P_{m}, \Theta_{np} ] &= i \hbar ( \delta_{mp} P_{n} - \delta_{mn} P_{p} )
\\
[\Theta_{mn}, \Theta_{pq} ] = i \hbar ( \delta_{mp} \Theta_{nq} &+ \delta_{nq} \Theta_{mp} - \delta_{np} \Theta_{mq} - \delta_{mq} \Theta_{np} )
\\
[\mathrm{h}, \Theta_{mn} ] &= 0~.
\end{align}
%5.8, 5.9, 5.10, 5.11

It is very interesting to note that the above algebra in contrast to the Heisenberg algebra (see
\cite{Singh:2018qzk})
%[83 palia])
admits finite-dimensional matrices to represent the operators $X_m$, $P_m$ and $\Theta_{mn}$ and therefore
the spacetime obtained above is a finite quantum system. Then clearly the above fuzzy
$dS_4$ falls into the general class of the fuzzy covariant spaces
\cite{Heckman:2014xha,Buric:2017yes,Barut}.
%[66, 67, 84 palia arithmisi].

\subsection{Gravity as Gauge Theory on the Fuzzy $dS_4$}
In the previous section, the fuzzy $dS_4$ space was constructed and the appropriate symmetry group to be used was found to be
the $SO(6)$. Following the recipe of the construction of Einstein gravity as gauge theory in
section \ref{subsec:4-d-Einstein},
in which the isometry group (the Poincar\'e group) was chosen to be gauged,
in this case the gauge group would be given by the isometry group of the fuzzy $dS_4$ space, namely the $SO(5)$,
viewed as a subgroup of the $SO(6)$ group.

However, it is known that in noncommutative gauge theories, the use of the anticommutators of the
generators of the algebra is inevitable, as we have explained in detail in our previous works
\cite{Chatzistavrakidis:2018vfi,Manolakos:2018isw}
%[42-43]
(see also
\cite{Aschieri2}).
%[15]) (oles me ti nea arithmisi).
Specifically, the anticommutation relations of the generators of the gauge group, $SO(5)$, produce operators that,
in general, do not belong to the algebra. The indicated treatment is to fix the representation of the generators
and all operators produced by the anticommutators of the generators to be included into the algebra, identifying
them as generators, too. This procedure led us  to an extension of the $SO(5)$ to $SO(6)\times U(1)$
($\sim U(4)$) group
\footnote{
Most probably the extension of the gauge group from $SO(5)$ to $SO(6)$ is not a coincidence, while the inclusion of
a $U(1)$ is quite intrinsic property of noncommutative theories.
}
with  the generators being represented by $4\times 4$ matrices in the spinor representation of $SO(6)$
(or the fundamental of $SU(4)$), 4.

In order to obtain the specific expressions of the matrices representing the generators, the four Euclidean $\Gamma$-matrices are employed, satisfying the following anticommutation relation:
\begin{equation}
\{\Gamma_{a}, \Gamma_{b} \} = 2 \delta_{ab} \one\,,
\end{equation}
%(5.12)
where $a,b = 1,\ldots ,4$. Also the $\Gamma_{5}$ matrix is defined as $ \Gamma_{5} = \Gamma_{1} \Gamma_{2} \Gamma_{3} \Gamma_{4} $. Therefore, the generators of the SO(6)$\times$U(1) gauge group are identified as:\\\\
a) Six generators of the Lorentz transformations:
$ \mathrm{M}_{ab} = - \tfrac{i}{4} [\Gamma_{a} , \Gamma_{b} ] = - \tfrac{i}{2} \Gamma_{a} \Gamma_{b}\,,a < b$,\\\\
b) four generators of the conformal boosts: $ \mathrm{K}_{a} = \tfrac{1}{2} \Gamma_{a}$,\\\\
c) four generators of the local translations: $ \mathrm{P}_{a} = -\tfrac{i}{2} \Gamma_{a} \Gamma_{5}$,\\\\
d) one generator for special conformal transformations: $\mathrm{D} = -\tfrac{1}{2} \Gamma_{5}$ and\\\\
e) one U(1) generator: $\one$. \\\\
The $\Gamma$-matrices are determined as tensor products of the Pauli matrices, specifically:
$$ \Gamma_{1} = \sigma_{1} \otimes \sigma_{1}, \ \ \ \Gamma_{2} = \sigma_{1} \otimes \sigma_{2}, \ \ \ \Gamma_{3} = \sigma_{1} \otimes \sigma_{3} $$
$$ \Gamma_{4} = \sigma_{2} \otimes \one, \ \ \  \Gamma_{5} = \sigma_{3} \otimes \one\,. $$
Therefore, the generators of the algebra are represented by the following 4$\times$4 matrices:
\begin{equation}
M_{ij} = - \frac{i}{2}\Gamma_{i} \Gamma_{j} = \frac{1}{2} \one \otimes \sigma_{k}\,,
\end{equation}
where $i,j,k = 1,2,3$ and:
\begin{equation}
M_{4k} = - \frac{i}{2}\Gamma_{4} \Gamma_{k} = - \frac{1}{2} \sigma_{3} \otimes \sigma_{k}\,.
\end{equation}
Straightforward calculations lead to the following commutation relations, which the operators satisfy:
\begin{equation}%5.15
\label{algebra}
\begin{split}
[ K_{a} , K_{b} ]& = i M_{ab}, \ \ \ [P_{a}, P_{b} ] = i M_{ab}  \\
[ X_{a}, P_{b} ]&  = i \delta_{ab}D, \ \ \ [X_{a}, D ] = i P_{a}  \\
[P_{a}, D ]& =i K_{a} , \ \ \ [K_a,P_b]=i\delta_{ab}D , \ \ \ [K_a,D]=-iP_a  \\
[K_{a}, M_{bc} ] &= i( \delta_{ac} K_{b} - \delta_{ab} K_{c} )  \\
[P_{a}, M_{bc} ] &= i( \delta_{ac} P_{b} - \delta_{ab} P_{c} )  \\
[M_{ab}, M_{cd} ]& = i( \delta_{ac} M_{bd} + \delta_{bd} M_{ac} - \delta_{bc}M_{ad} - \delta_{ad}M_{bc} )  \\
[D, M_{ab} ]& = 0\,.
\end{split}
\end{equation}

Having determined the commutation relations of the generators of the algebra, the noncommutative
gauging procedure can be done in a rather straightforward way. To start with, the covariant coordinate is defined as:
\begin{equation}%5.16
\label{covcoorddef}
\hat{X}_m=X_m\otimes \one +A_m(X)\,.
\end{equation}
The coordinate $\hat{X}_m$ is covariant by construction and this property is expressed as:
\begin{equation}%5.17
\label{covcoord}
\delta\hat{X}_m=i[\epsilon,\hat{X}_m]\,,
\end{equation}
where $\epsilon(X)$ is the gauge transformation parameter, which is a function of the coordinates
($N\times N$ matrices), $X_m$, but also is valued in the $SO(6)\times U(1)$ algebra. Therefore,
it can be decomposed on the sixteen generators of the algebra:
\begin{equation}%5.18
\label{epsilon}
\epsilon=\epsilon_0(X)\otimes \one +\xi^a(X)\otimes K_a+\tilde{\epsilon}_0(X)\otimes D+\lambda_{ab}(X)\otimes\Sigma^{ab}+\tilde{\xi}^a(X)\otimes P_a\,.
\end{equation}
Taking into account that a gauge transformation acts trivially on the coordinate $X_m$, namely $\delta X_m = 0$,
the transformation property of the $A_m$ is obtained by combining the equations (\ref{covcoorddef}), (\ref{covcoord}) and (\ref{epsilon}).
According to the corresponding procedure in the commutative case, the $A_m$ transforms in such a way
that admits the interpretation of the connection of the gauge theory. Also similarly to the case of
the gauge transformation parameter, $\epsilon$, the $A_m$, is a function of the coordinates $X_m$ of the fuzzy
space $dS_4$, but also takes values in the $SO(6)\times U(1)$ algebra, which means that it can be expanded on its
sixteen generators as follows:
\begin{equation}%5.19
\begin{split}
A_m(X)&=e_m^{~a}(X)\otimes P_a+\omega_{m}^{~ab}(X) \otimes \Sigma_{ab}(X) + b_{m}^{~a}(X) \otimes K_{a}(X) \\
&\qquad\qquad +\tilde{a}_{m}(X) \otimes D + a_{m}(X) \otimes  \one \,,
 \end{split}
\end{equation}
where it is clear that the various gauge fields have been corresponded to the generators of the $SO(6)\times U(1)$.
The component gauge fields are functions of the coordinates of the space, $X_m$, therefore they have the form
of $N\times N$ matrices, where $N$ is the dimension of the representation in which the coordinates are accommodated.
Thus, instead of the ordinary product, between the gauge fields and their corresponding generators,
the tensor product is used, since the factors are matrices of different dimensions, given that the generators
are represented by $4\times 4$ matrices. Then, each term in the expression of the gauge connection is
a $4N\times 4N$ matrix.

After the introduction of the gauge fields, the covariant coordinate is written as:
\begin{equation}%5.20
\hat{X}_{m} = X_{m} \otimes \one + e_{m}^{~a}(X) \otimes P_{a} + \omega_{m}^{~ab}(X) \otimes \Sigma_{ab} + b_{m}^{~a} \otimes K_{a} + \tilde{a}_{m} \otimes D + a_{m} \otimes \one \,.
\end{equation}
Then the next step in the theory that we are developing is to calculate its field strength tensor.
We found that for the fuzzy de Sitter space, the field strength tensor has to be defined as:
\begin{equation}%5.21
\label{fieldstrengtt}
\mathcal{R}_{mn} = [\hat{X}_{m}, \hat{X}_{n}]  - \frac{i\lambda^2}{\hbar}\hat{\Theta}_{mn}\,,
\end{equation}
where $\hat{\Theta}_{mn}=\Theta_{mn}\otimes \one +\mathcal{B}_{mn}$. The $\mathcal{B}_{mn}$ is a 2-form gauge field, which takes values in the
SO(6)$\times$U(1) algebra. The $\mathcal{B}_{mn}$ field was introduced in order to make the field strength tensor covariant, since in its absence
it does not transform covariantly
\footnote{Details on this generic issue on such spaces are given in Appendix A of the first paper of  \cite{Manolakos:2019fle}.}.
The $\mathcal{B}_{mn}$ field will contribute in the total action of the theory with a kinetic term of the following form:
\begin{equation}
\mathcal{S}_{\mathcal{B}}=\text{Tr}\,\text{tr}\, \hat{\mathcal{H}}_{mnp}\hat{\mathcal{H}}^{mnp}~.
\end{equation}
The $\hat{\mathcal{H}}_{mnp}$ field strength tensor transforms covariantly under a gauge transformation, therefore the above action is gauge invariant.

The field strength tensor of the gauge connection, \eqref{fieldstrengtt}, can be expanded in terms of the component curvature tensors, since it is valued in the algebra:
\begin{equation}
\begin{split}
\mathcal{R}_{mn}(X)& = R_{mn}^{~~~ab}(X) \otimes \Sigma_{ab} + \tilde{R}_{mn}^{~~a}(X) \otimes P_{a} + R_{mn}^{~~a}(X) \otimes K_{a} \\ 
&\qquad\qquad +\tilde{R}_{mn}(X) \otimes D + R_{mn}(X) \otimes \one \,.
\end{split}
\end{equation}
All necessary information for the determination of the transformations of the gauge fields and the expressions of the component curvature tensors is obtained.
The explicit expressions and calculations can be found in the first paper of ref.\cite{Manolakos:2019fle}.

\subsection{The Action and the Constraints for the Symmetry Breaking}
Concerning the action of the theory, it is natural to consider one of Yang-Mills type
\footnote{A Yang-Mills action $\text{tr}F^2$ defined on the fuzzy $dS_4$ space is gauge invariant,
for details see Appendix A of the first paper of \cite{Manolakos:2019fle}.}:
\begin{equation}%5.30
\mathcal{S}=\text{Tr}\text{tr}\{ \mathcal{R}_{mn},\mathcal{R}_{rs}\}\epsilon^{mnrs}\, ,
\end{equation}
where $\text{Tr}$ denotes the trace over the coordinates-$N\times N$ matrices (it replaces the integration of the
continuous case) and $\text{tr}$ denotes the trace over the generators of the algebra.

However the gauge symmetry of the resulting theory, with which we would like to end up, is the one described
by the Lorentz group, in the Euclidean signature, the $SO(4)$. In this direction, one could consider directly
a constrained theory in which the only component curvature tensors that would not be imposed to vanish would
be the ones that correspond to the Lorentz and the $U(1)$ generators of the algebra, achieving a breaking
of the initial $SO(6)\times U(1)$ symmetry to the $SO(4)\times U(1)$.
However, counting the degrees of freedom, adopting the above breaking would lead to an overconstrained theory.
Therefore, it is more efficient to follow a different procedure and perform the symmetry breaking in a
less straightforward way \cite{Manolakos:2019fle}. Accordingly, the first constraint is the
torsionless condition:
\begin{equation}%5.24
\label{ncconstraints}
\tilde{R}_{mn}^{~a}(P)=0\,,
\end{equation}
which is also imposed in the cases in which the Einstein and conformal gravity theories are described
as gauge theories. The presence of the gauge field $b_m^{~a}$ would admit an interpretation of a second
vielbein of the theory, that would lead to a bimetric theory, which is not what we are after in the
present case. Here it would be preferable to have the relation $e_m^{~a}={b}_m^{~a}$ in the solution of the
constraint. This choice leads also in expressing of the spin connection $\omega_m^{~ab}$ in terms of the rest
of the independent fields, ${e}_m^{~a}, {a}_m, {\tilde{a}}_m$. To obtain the explicit expression of the spin connection
in terms of the other fields, the following two identities are employed:
\begin{equation}%5.25
\label{ids}
\delta^{abc}_{fgh}=\epsilon^{abcd}\epsilon_{fghd}\quad\quad\text{and}\quad\quad \frac{1}{3!}\delta^{abc}_{fgh}a^{fgh}=a^{[fgh]}\,.
\end{equation}
Solving the constraint $\tilde{R}(P)=0$, it follows that:
\begin{equation}%5.26
\epsilon^{abcd}[e_m^{~b},\omega_n^{~cd}]-i\{\omega_m^{~ab},e_{nb}\}=-[D_m,e_m^{~a}]-i\{e_m^{~a},\tilde{a}_m\}\, ,
\end{equation}
where $D_m = X_m +a_m$ being the covariant coordinate of an Abelian noncommutative gauge theory.
Then the above equation leads to the following two:
\begin{equation}%5.27
\epsilon^{abcd}[e_m^{~b},\omega_n^{~cd}]=-[D_m,e_m^{~a}]\quad\quad \text{and}\quad\quad \{\omega_m^{~ab},e_{nb}\}=\{e_m^{~a},\tilde{a}_n\}\,.
\end{equation}
Taking into account also the identities, (\ref{ids}), the above equations lead to the desired expression
for the spin connection in terms of the rest fields:
\begin{equation}%5.28
\label{omegaintermsofe}
\omega_n^{~ac}=-\frac{3}{4}e^m_{~b}(-\epsilon^{abcd}[D_m,e_{nd}]+\delta^{[bc}\{e_n^{~a]},\tilde{a}_m\})\,.
\end{equation}
According to
\cite{Green:1987mn},
%[87 palia arithmisi],
the vanishing of the field strength tensor in a gauge theory could
lead to the vanishing of the associated gauge field.  However, the vanishing of the torsion component
tensor, $\tilde{R}(P)=0$, does not imply $e_\mu^{~a}=0$, because such a choice would lead to degeneracy of the metric
tensor of the space
\cite{Witten:1988hc}.
%[12].
The field that can be gauge-fixed to zero is the $\tilde{a}_m$. Then this fixing, $\tilde{a}_m=0$,
will modify the expression of the spin connection, (\ref{omegaintermsofe}), leading to a further
simplified expression of the spin connection in terms of the vielbein:
\begin{equation}%5.29
\label{omegaintermsofeteliko}
\omega_n^{~ac}=\frac{3}{4}e^m_{~b}\epsilon^{abcd}[D_m,e_{nd}]\,.
\end{equation}
We note that the $U(1)$ field strength tensor, $R_{mn}(\one)$, signaling the noncommutativity of the space,
is not considered to be vanishing. The $U(1)$ remains unbroken in the resulting theory after the breaking,
since we still have a theory on a noncommutative space. However, the corresponding field, $a_m$, would vanish
if we consider the commutative limit of the broken theory, in which noncommutativity is lifted and
$a_m$ decouples being super heavy. In this limit, the gauge theory would be just $SO(4)$.
Alternatively, another way to break the $SO(6)$ gauge symmetry to the desired $SO(4)$ is to induce
a spontaneous symmetry breaking by including two scalar fields in the 6 representation of $SO(6)$
\cite{Li:1973mq},
extending the argument developed for the case of the conformal gravity to the noncommutative framework.
It is expected that the spontaneous symmetry breaking induced by the scalars would lead to a
constrained theory as the one that was obtained above by the imposition of the constraints (\ref{ncconstraints}).
After the symmetry breaking, i.e. including the constraints, the surviving terms of the action will be:
\begin{equation}%5.31
\begin{split}
\mathcal{S}&=2\text{Tr}(R_{mn}^{~~~ab}R_{rs}^{~~cd}\epsilon_{abcd}\epsilon^{mnrs}+4\tilde{R}_{mn}R_{rs}\epsilon^{mnrs}\\
&\qquad\qquad +\frac{1}{3}H_{mnp}^{~~~~ab}H^{mnpcd}\epsilon_{abcd}+\frac{4}{3}\tilde{H}_{mnp}H^{mnp})\,.
\end{split}
\end{equation}

Finally replacing with the explicit expressions of the component tensors and writing the $\omega$ gauge field
in terms of the surviving gauge fields, (\ref{omegaintermsofeteliko}) and then varying with respect to
the independent gauge fields would lead to the equations of motion.

\section{Summary and Conclusions}
In the present review we presented a 4-d gravity model as a gauge theory on a fuzzy version of the 4-d de Sitter space.
It should be stressed that the constructed fuzzy $dS_4$ consists a 4-d covariant noncommutative space,
respecting Lorentz invariance, which is of major importance in our case.
Next, although we started by gauging the isometry group of $dS_4$, $SO(5)$, we were led to enlarge it to $SO(6)\times U(1)$
in order to include the anticommutators of its generators that appear naturally in the noncommutative framework
and in fixing the respresentation. Then, following the standard procedure we calculated the transformations
of the fields and the expressions of the component curvature tensors. Since our aim was to result with a theory
respecting the Lorentz symmetry, we imposed certain constraints in order to break the initial symmetry.
After the symmetry breaking, the action takes its final form and its
variation will lead to the equations of motion. The latter will be part of our future work. It should be noted that,
before the symmetry breaking, the results of the above construction reduce to the ones of the conformal gravity
in the commutative limit. Finally, it should be also emphasized that the above is a matrix model giving insight
into the gravitational interaction in the high-energy regime and also giving promises for improved UV properties
as compared to ordinary gravity. Clearly, the latter, as well the inclusion of matter fields is going to be a subject of further study.
\\

Acknowledgements:
We would like to thank Ali Chamseddine, Paolo Aschieri, Thanassis Chatzistavrakidis, Evgeny Ivanov, Larisa Jonke, Danijel Jurman, Alexander Kehagias, Dieter Lust, Denjoe O'Connor, Emmanuel Saridakis, Harold Steinacker, Kelly Stelle, Patrizia Vitale and Christof Wetterich for
useful discussions. The work of two of us (GM and GZ) was partially supported by the COST Action MP1405, while both would like to thank ESI - Vienna for the hospitality during their participation in the Workshop ``Matrix Models for Noncommutative Geometry and String Theory'', Jul 09 - 13, 2018. One of us (GZ) has been supported within the Excellence Initiative funded by the German and States Governments, at the Institute for Theoretical Physics, Heidelberg University and from the Excellent Grant Enigmass of LAPTh. GZ would like to thank the ITP - Heidelberg, LAPTh - Annecy and MPI - Munich for their hospitality.\\
Last but not least GZ thanks the organisers of the Workshop in Varna for their warm hospitality.

\end{document}